\documentclass[aps,prb,showpacs,reprint, superscriptaddress]{revtex4-1}

\usepackage{graphicx}
\usepackage{color}
\usepackage{amsmath}
\usepackage{bbm}

\input epsf

\begin{document}

\title{Intersubband pairing induced Fulde-Ferrell phase in metallic nanofilms}
\author{P. W\'ojcik}
\email{pawel.wojcik@fis.agh.edu.pl}
\affiliation{AGH University of Science and Technology, Faculty of
Physics and Applied Computer Science, 30-059 Krakow, Poland
Al. Mickiewicza 30, 30-059 Krakow, Poland}

\author{M. P. Nowak}
\affiliation{AGH University of Science and Technology, Academic Centre
for Materials and Nanotechnology, Al. Mickiewicza 30, 30-059 Krakow,
Poland}

\author{M. Zegrodnik}
\affiliation{AGH University of Science and Technology, Academic Centre
for Materials and Nanotechnology, Al. Mickiewicza 30, 30-059 Krakow,
Poland}

\begin{abstract}
We consider a free-standing metallic nanofilm with a predominant intersubband paring which emerges as a result of the confinement in the growth direction. We show that the Fermi wave vector mismatch between the subbands, detrimental to the intersubband pairing, can be compensated by the non-zero center of mass momentum of the Cooper pairs.  This leads to the spontaneous appearance of the intersubband Fulde-Ferrell (IFF) state, even in the absence of an external magnetic field. Our study of the intrasubband pairing channel on the stability of the IFF phase shows that the former strongly competes with the intersubband pairing, which prohibits the coexistence of the two superconducting phases. Interestingly, upon application of the magnetic field we find a transition to an exotic mixed spin-singlet subband-triplet and spin-triplet subband-singlet paired state. Finally, we discuss the possibility of existence of the IFF pairing in novel superconducting materials.
\end{abstract}

\maketitle
\section{Introduction}

In the last decade, the study of superconductors in the nanoscale regime has evolved into one of the most active research directions in the solid state physics. This was mostly driven by the rapid progress in growth and characterization techniques which allow metallic films, and other types of superconducting materials, to be fabricated with atomic precision\cite{Zhang2010,Uchihashi2011,Qin2009}. In metallic nanofilms, due to the strong confinement of electrons in the growth direction, the Fermi surface splits into series of subbands what results in multiband superconductivity similar to that observed in superconductors such as, e.g. MgB$_2$\cite{Souma2003,Giubileo2001,Chen2001,Szabo2001} or iron-pnictides~\cite{Kuroki2008, Dai2008, Jeglic2010, Singh2008}. The multiband character of superconductivity in metallic nanofilms was confirmed by measurements of the critical temperature\cite{Ozer2006,Eom2006,Guo2004} and magnetic field\cite{Bao2005,Sekihara2013} oscillations as a function of the nanofilm thickness. This behavior was explained as resulting from the van Hove sigularities occurring each time when the bottom of the subband passes through the Fermi level (Lifshitz transition)\cite{Shanenko2008,Shanenko2007,Wojcik2014}.

When the material becomes thin enough, the confinement affects not only the electronic spectrum, but also the phononic degrees of freedom. The phonon dispersion in thin films strongly deviates from that observed in the bulk\cite{Nabity1991} what changes the electron-phonon coupling. Determination of the electron-phonon coupling in individual subbands and between them in metallic nanofilms is still an open issue, from both theoretical and experimental point of view. In particular, when the energy between electronic states (subbands) becomes smaller than the energy of the electron-phonon interaction\cite{Shanenko2015}, the unconventional intersubband pairing can appear.  In this paper, we show  that in metallic nanofilms, the existence of such an exotic intersubband pairing can spontaneously induce a superconducting phase with a finite center-of-mass momentum of the Cooper pairs. 

The finite momentum pairing was originally introduced in the 60's by Fulde and Ferrel~\cite{Fulde1964} as well as Larkin and Ovchinnikov~\cite{Larkin1964} as resulting from the paramagnetic effect.  In the external magnetic field the Zeeman splitting of the Fermi surface generates the Fermi wave vector mismatch between spin-up and spin-down electrons which is detrimental for the paring of electrons in spin-singlet state. The Fermi wave vector mismatch can be compensated by the non-zero center-of-mass momentum of the Copper pairs leading to the Fulde-Ferrell (FF) phase\cite{Maska2010,Ptok2013_1,Ptok2014_1,Takahashi2014}. Although the physics standing behind the finite momentum Cooper pairing is relatively transparent, its experimental observation turned out to be extremely challenging. This is mainly due to the dominant role of the orbital effects which suppress the critical field well below the range of the FF phase stability\cite{Gruenberg1966}. For this reason, the existence of this unconventional paired state has been proposed to appear in 2D organic superconductors\cite{Singleton2000,Tanatar2002,Uji2006,Shinagawa2007} or ultra-thin films\cite{Wojcik2016,Croitoru2012,Croitoru2012b} subject to an in-plane magnetic field, where the orbital effects are strongly suppressed by the confinement in the growth direction. Alternatively, the FF phase is believed to appear in heavy fermion systems\cite{Bianchi2003,Kumagai2006,Correa2007,Matsuda2007} where the orbital effects are suppressed by the high effective mass or superconducting nanowires\cite{Wojcik2015,Mika_2017}. So far, in the ongoing debate on the stability of the FF phase\cite{Wang2018}, the strong experimental evidence of the finite-momentum pairing has been provided only in 2D organic superconductors\cite{Uji2006,Koutroulakis2016,Beyer2012,Mayaffre2014} and superconductor/ferromagnet\cite{Buzdin2005} (topological insulator\cite{Hart2017,Chen2018}) junctions for which  the FF phase is formed in the proximitized part of the junction.

In this paper, we present that the FF phase can appear spontaneously (without the magnetic field) in superconducting metallic nanofilms as a result of the quantization of the electronic bands in the growth direction. When the energy between quantized electronic states is smaller than the energy of the electron-phonon interaction the electrons from different bands can create Cooper pairs. Then, the Fermi wave vector mismatch between the subbands is compensated by the non-zero center-of-mass momentum of the Cooper pairs leading to the spontaneous FF phase formation induced by the intersubband pairing.


The structure of the paper is as follows. In section \ref{sec:2} we introduce the theoretical model of the non-zero momentum paring in metallic nanofilms. In section \ref{sec:3} we present our results both in the absence and in the presence of the magnetic field. Conclusions and outlook are provided in section \ref{sec:4}.

\section{Theoretical model}
\label{sec:2}
\subsection{FF phase in a metallic nanofilm}

\begin{figure}[ht]
\begin{center}
\includegraphics[scale=0.5]{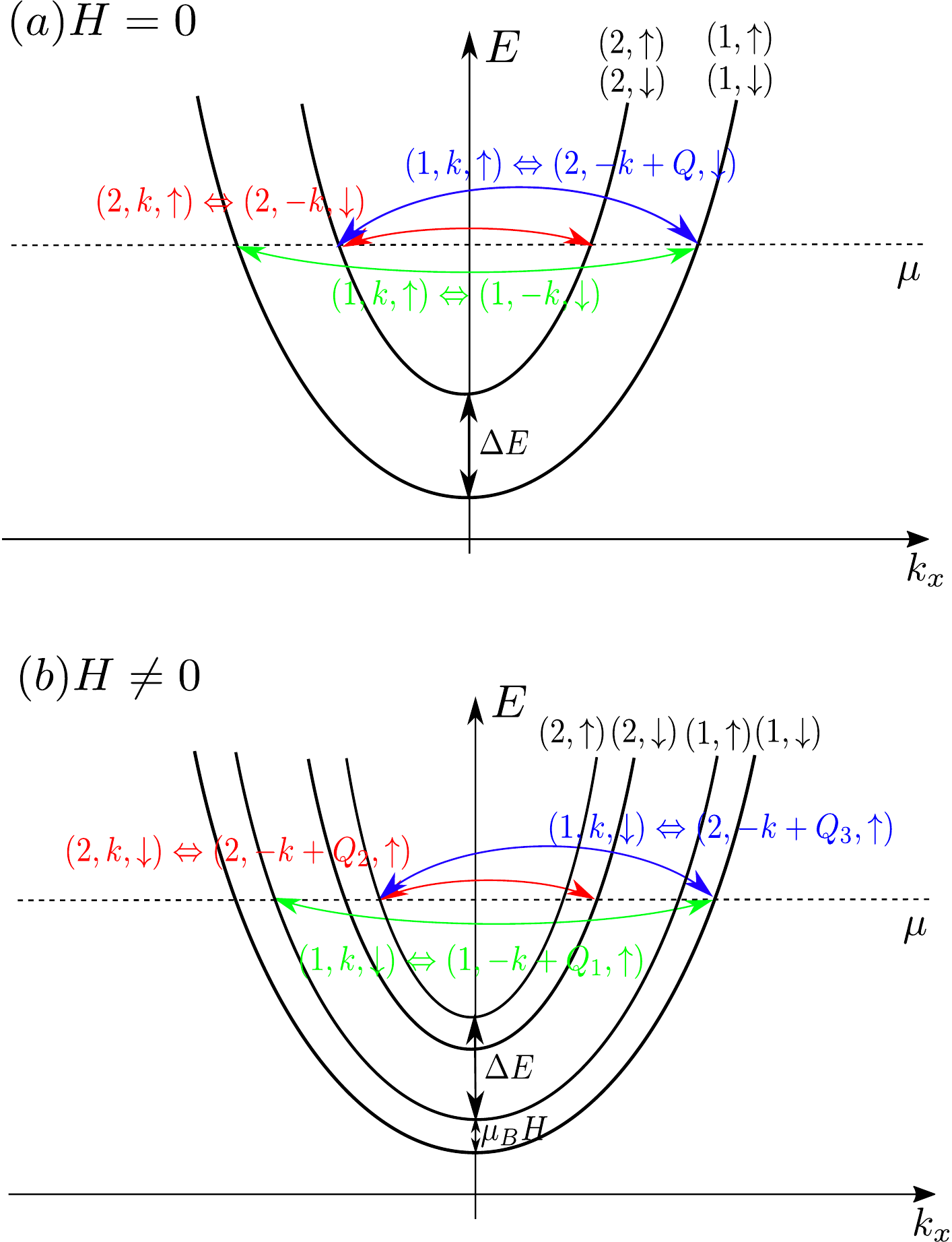}
\caption{Schematic illustration of pairing in the two-band model (cross section along $k_y=0$) for the magnetic field (a) $H=0$ and (b) $H\ne0$. Electronic states from opposite sides of the Fermi surface, forming Cooper pairs, are connected by arrows. The intersubband finite momentum pairing is marked by the blue arrows. The horizontal dashed line denotes the Fermi level, $\mu$.}
\label{fig1}
\end{center}
\end{figure}

We consider Pb(111) free-standing nanofilms with predominant intersubband electron-phonon coupling. The first-principle calculations\cite{Wei2002} of the quantized band structure for Pb nanofilms showed that the quantum size effect in (111) direction can be well described only by the quantum well states centered at the L point of the Brillouin zone, where the energy dispersion is nearly parabolic. Based on those results, here we use the parabolic band approximation. For simplicity, we consider two lowest subbands assuming that they are separated by the energy $\Delta E$. The two band model allows to underline the sole role of the intersubband paring without any disturbances resulting from Cooper pairs tunnelling to other subbands. The schematic illustration of paring in the considered model is presented in Fig.~\ref{fig1}.

For $\mathbf{H}=0$ and small intersubband coupling strength, the superconducting state is crated as a result of pairing between electrons with opposite spins and momenta within the subbands, $(1,\mathbf{k},\uparrow) \Leftrightarrow (1,-\mathbf{k},\downarrow)$ and $(2,\mathbf{k},\uparrow) \Leftrightarrow (2,-\mathbf{k},\downarrow)$. If we increase the intersubband coupling, a new paring  channel opens, where electrons from different subbands form the Cooper pairs $(1,\mathbf{k},\uparrow) \Leftrightarrow (2,-\mathbf{k}+\mathbf{Q},\downarrow)$. In such a case, the wave vector mismatch between electrons with opposite spins is compensated by the non-zero momentum of the pairs, $\mathbf{Q}$, leading to the spontaneous (without magnetic field) FF phase [Fig.~\ref{fig1}(a)], in which not all of the particles at the Fermi surface are paired. The further increase of the intersubband coupling above a critical value leads to a situation in which the pairing region in reciprocal space is relatively large meaning that the Fermi wave vector mismatch cannot prevent the electrons from pairing. In such a scenario, the electrons from the more populated subband are moved to the less populated one and the intersubband pairing with zero momentum $\mathbf{Q}=0$ is more preferable.

If we apply the magnetic field $\mathbf{H}\ne0$ [Fig.~\ref{fig1}(b)] both the subbands split as a result of Zeeman effect. The wave vector mismatch between electrons with opposite spins appears separately in each subband which may result in the finite momentum pairing in each of them. In general, the Cooper pair momentum $\mathbf{Q}$ for both the subbands and between them can be different leading to  the FF phase which is a superposition of phases with three different $\mathbf{Q}$ vectors (periods of the energy gap oscillations in the real space). In this case, the appearance of the intra and intersubband FF phase as well as the coexistence of them is determined by the spin-splitting energy and the energy separation $\Delta E$. In particular, the Zeeman splitting can compensate the energy $\Delta E$ between the states $(1,\uparrow)$ and $(2,\downarrow)$ making the paring between them more preferable while the paring between the spin reversed states $(1,\downarrow)$ and $(2,\uparrow)$ is destroyed by the large vector mismatch between the states at the Fermi level, which increases with the increasing magnetic field.

\subsection{Two-band model with intersubband paring}
\begin{figure*}[ht]
\begin{center}
\includegraphics[scale=0.4]{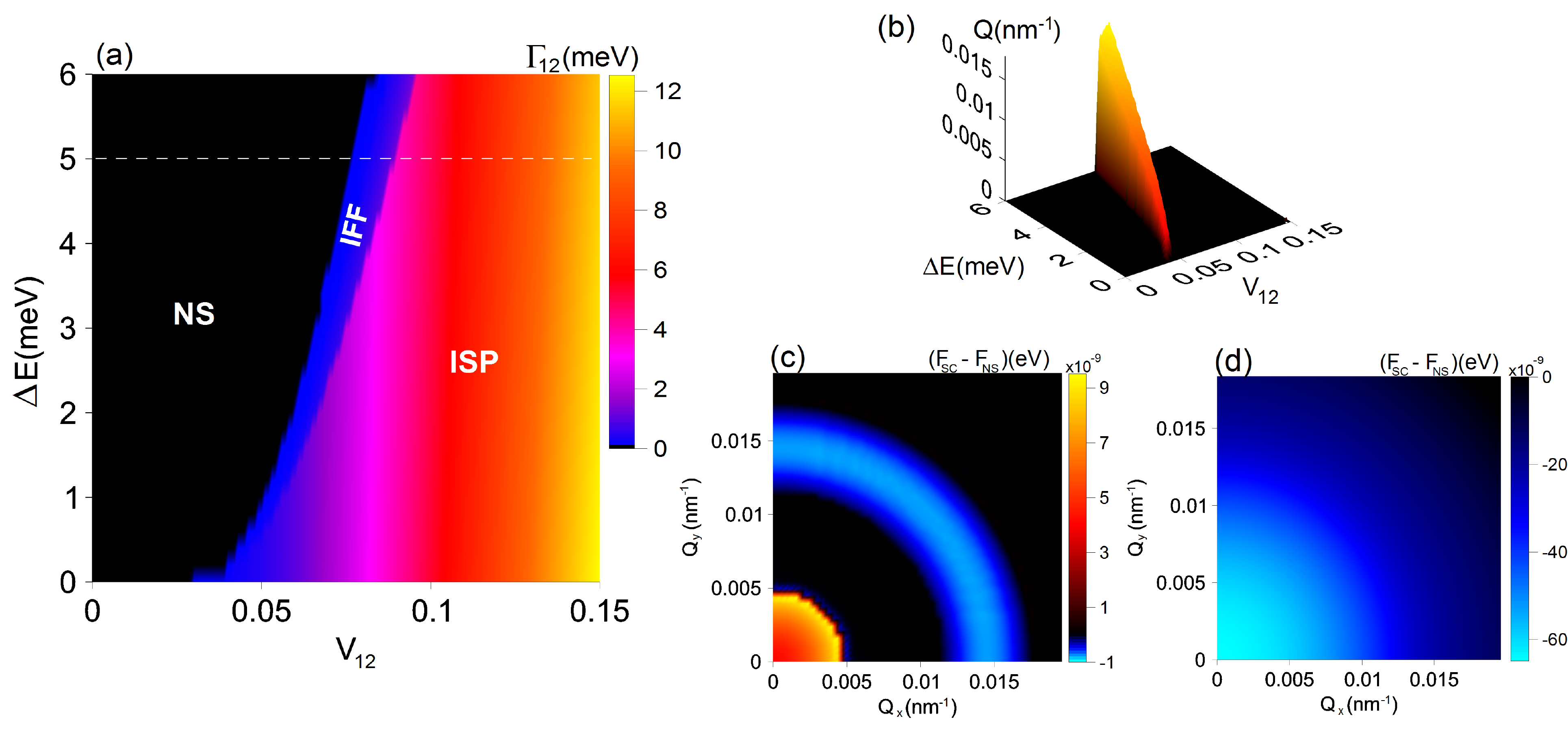}
\caption{(a) The phase diagram in $(V_{12},\Delta E)$ plane, where one can distinguish the regions of NS (normal state), IFF (intersubband Fulde-Ferrell phase), and  ISP (intersubband paired phase). The dashed horizontal line denotes $\Delta E= 5$~meV chosen for the further analysis; (b) the Cooper-pair momentum, which minimizes the energy of the system, as a function of both $\Delta E$ and $V_{12}$; (c) and (d) the free energy of the system as a function of the Cooper pair momentum $\mathbf{Q}=(Q_x,Q_y)$ for two selected values of the interband pairing strength $V_{12}$, for which the IFF (c) and ISP (d) phases are stable, respectively.}
\label{fig2}
\end{center}
\end{figure*}
We start from the general form of the mean-field-BCS Hamiltonian in the presence of external in-plane magnetic field
$\mathbf{H}=(H,0,0)$
\begin{eqnarray}
\hat{\mathcal{H}}&=&\sum _{\sigma} \int d^3 r
\hat{\Psi}^{\dagger} (\mathbf{r},\sigma) \hat{H}_e ^{\sigma}
\hat{\Psi}(\mathbf{r},\sigma) \nonumber \\ 
&+& \int d^3 r \left [ \Delta
(\mathbf{r})\hat{\Psi}^{\dagger}(\mathbf{r},\uparrow)
\hat{\Psi}^{\dagger}(\mathbf{r},\downarrow) +H.c. \right ] \nonumber \\
&+&\int d^3r \frac{|\Delta(\mathbf{r})|^2}{g},
\label{eq:ham}
\end{eqnarray}
where $\sigma$ corresponds to the spin state $(\uparrow, \downarrow)$, $g$ is the phonon-mediated electron-electron coupling constant, $\hat{H}_e^{\sigma}$ is the single-electron Hamiltonian and $\Delta(\mathbf{r})$ is the superconducting gap parameter in real space defined as
\begin{equation}
 \Delta(\mathbf{r})=-g \left < \hat{\Psi} (\mathbf{r},\downarrow)
\hat{\Psi} (\mathbf{r},\uparrow)  \right >.
\label{eq:gap_def}
\end{equation}

In the two band model the field operators have the form
\begin{eqnarray}
 \hat{\Psi}(\mathbf{r},\sigma)=\sum_{\mathbf{k}} \left (
\phi_{1\mathbf{k}}(\mathbf{r})\:\hat{c}_{1 \mathbf{k} \sigma} + \phi_{2\mathbf{k}}(\mathbf{r})\:\hat{c}_{2 \mathbf{k} \sigma} \right ) , \\
\hat{\Psi}^{\dagger}(\mathbf{r},\sigma)=\sum_{\mathbf{k}} \left (
\phi^*_{1 \mathbf{k}}(\mathbf{r})  \: \hat{c}^{\dagger}_{1 \mathbf{k} \sigma} + \phi^*_{2 \mathbf{k}}(\mathbf{r})  \: \hat{c}^{\dagger}_{2 \mathbf{k} \sigma} \right ),
\label{eq:field_op}
\end{eqnarray}
where $\hat{c}_{n \mathbf{k} \sigma}
(\hat{c}^{\dagger}_{n \mathbf{k} \sigma})$ with $n=(1,2)$ is
the anihilation (creation) operator for an electron with spin $\sigma$ in the subband $n$ characterized by
the wave vector $\mathbf{k}$ and $\phi_{n \mathbf{k}}(\mathbf{r})$ are the single-electron eigenfunctions  of the Hamiltonian $\hat{H}^{\sigma}_e$.

The FF phase, induced either by the intersubband coupling or the Zeeman splitting, is characterized by the pairing with the non-zero momentum $(\mathbf{k},\uparrow) \Leftrightarrow (-\mathbf{k}+\mathbf{Q}, \downarrow)$. The two band Hamiltonian with the finite momentum pairing takes the form 
\begin{equation}
\begin{split}
\hat{\mathcal{H}}^{\mathbf{Q}}&=
\sum_{\mathbf{k}}\mathbf{\hat{f}}^{\dagger}_{\mathbf{k},\mathbf{Q}} H_{\mathbf{k}}^{\mathbf{Q}}\mathbf{\hat{f}}_{\mathbf{k},\mathbf{Q}}+\sum_{\mathbf{k}} \left ( \xi_{1, -\mathbf{k}+\mathbf{Q},\downarrow} + \xi_{2,-\mathbf{k}+\mathbf{Q}, \downarrow} \right ) \\ 
& +\sum_{n,m=1,2} \frac{|\Delta_{n,m}^{\mathbf{Q}}|^2}{g},  
\end{split}
\label{eq:Ham_matrix_1}
\end{equation}
where $\mathbf{\hat{f}}^{\dagger}_{\mathbf{k},\mathbf{Q}}=(\hat{c}^{\dagger}_{1,\mathbf{k},\uparrow}, \hat{c}_{1,-\mathbf{k}+\mathbf{Q},\downarrow}, \hat{c}^{\dagger}_{2,\mathbf{k},\uparrow}, \hat{c}_{2,-\mathbf{k}+\mathbf{Q},\downarrow})$ is the composite vector operator and
\begin{equation}
H_{\mathbf{k}}^{\mathbf{Q}}=\left(\begin{array}{cccc}
\xi_{1,\mathbf{k},\uparrow} & \Gamma_{1,1}^{\mathbf{Q}} & 0 & \Gamma_{1,2}^{\mathbf{Q}}\\
\Gamma_{1,1}^{\mathbf{Q}} & -\xi_{1, -\mathbf{k}+\mathbf{Q}, \downarrow} & \Gamma_{2,1}^{\mathbf{Q}} & 0 \\
0 & \Gamma_{2,1}^{\mathbf{Q}} & \xi_{2,\mathbf{k},\uparrow} & \Gamma_{2,2}^{\mathbf{Q}} \\
\Gamma_{1,2}^{\mathbf{Q}} & 0 & \Gamma_{2,2}^{\mathbf{Q}} & -\xi_{2, -\mathbf{k}+\mathbf{Q}, \downarrow}
\end{array} \right).
\label{eq:matrix_H}
\end{equation}
In the above Hamiltonian, $\xi_{n,\mathbf{k}}$ are the single-particle energies which, in parabolic band approximation, are given by
\begin{equation}
\begin{split}
\xi_{1,\mathbf{k},\sigma}&=E_0+\frac{\hbar^2(k_x^2+k_y^2)}{2m}+ s \mu_B H - \mu, \\
\xi_{2,\mathbf{k}, \sigma}&=E_0+\frac{\hbar^2(k_x^2+k_y^2)}{2m}+s \mu_B H+\Delta E - \mu,
\end{split}
\end{equation}
where $m$ is the electron mass, $\mu$ is the Fermi energy, $s=\pm 1$ for the spin index $\sigma=(\uparrow,\downarrow)$, $\mu_B$ is the Bohr magneton, $\mathbf{k}=(k_x,k_y)$ and $E_0$ is the bottom of the lower subband, assumed to be the reference energy ($E_0=0$). Due to the strong confinement in the growth direction, the orbital effects from the in-plane magnetic field can be neglected.\\ 
The intra and intersubband superconducting gap parameters $\Gamma_{n,m}^{\mathbf{Q}}$ are expressed by 
\begin{equation}
 \Gamma_{n,m}^{\mathbf{Q}} = -g \sum _{n',m'=1,2} V_{n,m}^{n',m'} \Delta _{n',m'} ^{\mathbf{Q}}, 
 \label{eq:gamma}
\end{equation}
where the interaction matrix elements 
\begin{equation}
 V_{n,m}^{n',m'}=\int d^3r \phi^*_{n\mathbf{k}}(\mathbf{r}) \phi^*_{m\mathbf{k}}(\mathbf{r}) \phi_{n'\mathbf{k}}(\mathbf{r}) \phi_{m'\mathbf{k}}(\mathbf{r}).
\label{eq_inter}
\end{equation}
and 
\begin{equation}
 \Delta _{n,m} ^{\mathbf{Q}} = \sum_k{}^{'}  \langle \hat{c}_{n, \mathbf{-\mathbf{k}+\mathbf{Q}}, \downarrow} \hat{c}_{m, \mathbf{k}, \uparrow } \rangle.
 \label{eq:delta}
 \end{equation}
 The primed sum in (\ref{eq:delta}) means that the summation is carried out only if both the single electron states $\xi_{n,\mathbf{k}}$ and $\xi_{m,\mathbf{k}}$ are located inside the Debye window $[\mu-\hbar \omega _D, \mu + \hbar \omega _D]$, where $\omega _D$ is the Debye frequency.  \\
Note, that in the parabolic band approximation $V_{n,m}^{n',m'}$ does not depend on the $\mathbf{k}$ vector.  Moreover, due to the symmetry of the single-electron eigenfunctions (we assume the hard wall potential) $V_{n,n}^{n,m}=V_{n,n}^{m,n}=V_{n,m}^{m,m}=V_{n,m}^{n,n}=0$ and $V_{n,n}^{m,m}=V_{m,m}^{n,n}=V_{n,m}^{n,m}=V_{n,m}^{m,n}$. This in turn reduces the interaction matrix to three non-zero elements: two intrasubband coupling constants,  which we assumed to be equal $V=V^{11}_{11}=V^{22}_{22}$, and the intersubband coupling constant, $V_{12}=V_{11}^{22}$. The interaction matrix elements $V$ and $V_{12}$ determine the effective electron-electron interaction  by changing the $g$ parameter for the intra- and inter-subband paring. In the following part of the paper we treat $V$ and $V_{12}$ as dimensionless parameters which control the pairing by rescaling $g$ expressed in eV.

The numerical diagonalization of (\ref{eq:matrix_H}) leads to the quasiparticle energies which are then used to derive the free-energy functional in a standard statistical-mechanical manner. The paring energies $\Gamma _{n,m}^{\mathbf{Q}}$ are obtained by solving the set of self-consistent equations, while the $\mathbf{Q}$ vector is determined by minimizing the free energy of the system.

The calculations were carried out for material parameters corresponding to Pb: $m=1$, $gN_{bulk}=0.18$ where  $N_{bulk}=mk_F/(2\pi^2\hbar^2)$ is the bulk density of the single-electron states at the Fermi level, $\hbar \omega _D=32.31$~meV and $\mu \gg \Delta E, E_0$.

\section{Results}
\label{sec:3}

In the first part of our analysis, we focus on the intersubband pairing and the formation of spontaneous FF phase. For the sake of simplicity, we initially consider a situation with $V=0$ (no intrasubband pairing). The influence of the latter as well as the case of non-zero external magnetic field are analyzed in the following part of the paper. 
\begin{figure}[ht]
\begin{center}
\includegraphics[scale=0.35]{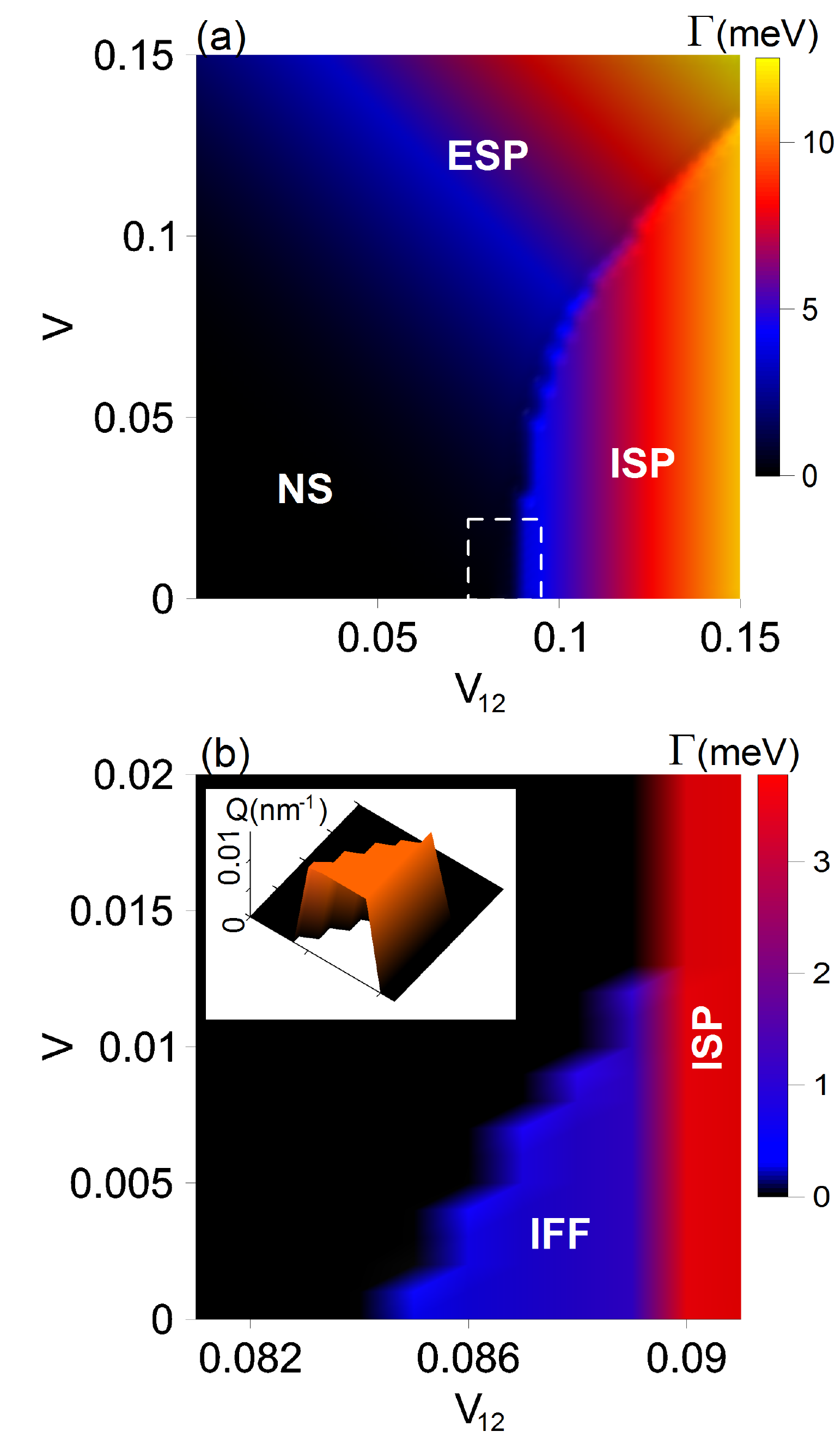}
\caption{(a) The superconducting gap parameter as a function of both inter- and intra-subband pairing strength with visible regions of stability of ISP (intersubband paired state), ESP (equal subband paired state), and NS (normal state). Results for $\mathbf{Q}=0$  and a selected value of the energy separation $\Delta E=5$~meV; (b) the region of stability of the Fulde-Ferrell inter-subband state with the values of $\mathbf{Q}$ vector which minimize the free energy in the inset. Results for $\mathbf{Q}\ne 0$ within the parameter range marked by the white rectangle in panel (a). }
\label{fig3}
\end{center}
\end{figure}
\begin{figure*}[t]
\begin{center}
\includegraphics[scale=0.35]{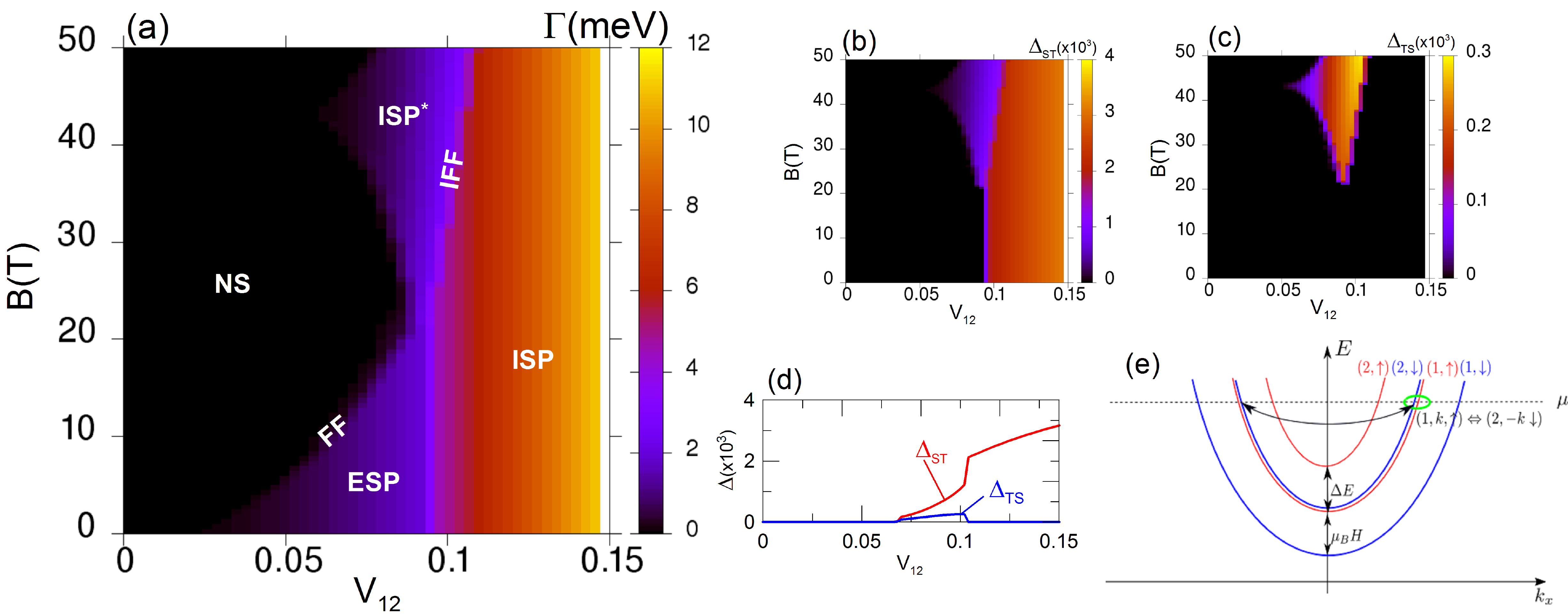}
\caption{(a) The phase diagram in the ($V_{12},B$) plane for selected value of the intra-subband pairing $V=0.05$; (b) the spin-singlet, subband-triplet pairing amplitude as well as (c) the spin-triplet, subband-singlet pairing amplitude \textit{vs.} $V_{12}$ and $B$. In (d) we show $\Delta_{ST}$ and $\Delta_{TS}$ as a function of $V_{12}$ for a selected value of $B$. Additionally in (e) the schematic representation of the spin-splitted subbands in the external magnetic field is provided.}
\label{fig4}
\end{center}
\end{figure*}

In Fig. \ref{fig2}(a) we show the phase diagram in the $(V_{12},\Delta E)$ plane, with the SC gap marked by the colors. As one can see, a significant stability regions of the normal state (NS) and intersubband superconducting phase (ISP) appear in the diagram.  The energy separation between the bands has a detrimental influence on the intersubband pairing as it introduces the Fermi wave vector mismatch between the paired electrons. 
Different energies of the band bottom with respect to the Fermi energy results in disproportion between the number of particles in each of them. In such a case the pairing  becomes energetically unfavorable since not all particles can be paired. Nevertheless, for non-zero  $\Delta E$ and high enough values of $V_{12}$ the width of the pairing region around the Fermi surfaces is relatively large, meaning that the Fermi wave vector mismatch no longer prevents the particles from forming the Cooper pairs. In such situation, the particles from the more populated subband are transferred to the less populated one, and no unpaired particles are left\cite{Moreo2009}. Those two effects lead to opening of the gap across the full Fermi surface. The resulting phase resembles a BCS superconductor, except here the Cooper pairs are formed by particles from two different subbands (ISP - intersubband paired state).

An interesting region lies in between NS and ISP, for which the intersubband pairing is already too small to induce a fully gapped paired state, however, superconductivity can still appear in the form of the intersubband Fulde-Ferell phase (IFF). In the latter, the Fermi wave vector mismatch between the two subbands is compensated by the non-zero center-of-mass momentum of the Cooper pairs, which leads to a non-homogeneous SC gap with a small depairing region in the Brillouin zone, occupied by the unpaired electrons. In Fig. \ref{fig2}(b) we show the values of the Cooper pair momenta which corresponds to the minimum of the free energy. Since we are working in the parabolic band approximation, the direction of the $\mathbf{Q}$ vector can be chosen arbitrary. This is clearly seen in Fig. \ref{fig2}(c) and (d), where the free energy of the system is plotted as a function of both $Q_x$ and $Q_y$, for two selected values of $V_{12}$. One should note, that the obtained rotational symmetry is also a result of the chosen pairing symmetry, which is assumed to be $s$-$wave$, in our case. In general, for other pairing symmetries such as e.g. $d$-$wave$ the rotational symmetry in the $(Q_x,Q_y)$ space can be broken\cite{Zegrodnik2015,Zegrodnik2014}. As can be seen, for the situation shown in (d) the nonzero Cooper pairing is not energetically favorable, since the energy of the system increases with increasing $Q$. On the other hand, the energy minimum for non-zero $Q$ shown in (c) signals the possibility of non-zero momentum pairing. In such situation the SC gap parameter changes its phase in real space as one moves along the direction determined by the modulation vector. The discontinuity in $\Gamma_{12}$ at the border between ISP and IFF indicates a first order phase transition between the two. Such a behavior is also reported for the case of the conventional Fulde-Ferrell phase induced by the presence of an external magnetic field\cite{Maska2010}. However, here the non-zero momentum pairing appear spontaneously between two subbands without the application of an external magnetic field.

Our simulations for non-zero temperatures (not shown) demonstrate that the stability region of the IFF phase gradually decreases with increasing temperature whereas the phase transition between the IFF phase and the ISP phase is non-continues, which is indicated by a sudden drop of the gap parameter $\Gamma_{12}$.

Next, we analyze the influence of the intrasubband pairing on the phase diagram analyzed so far. The situation corresponding to both $V_{12}\neq 0$ and $V\neq 0$ is shown in Fig.~\ref{fig2}, for $\Delta E=5$~meV (marked by the horizontal line in Fig.~\ref{fig2}(a)). Since the calculations including the non-zero momentum paring are highly time consuming we start from the simplified phase diagram with $\mathbf{Q}=0$. Fig.~\ref{fig3}(a) presents relatively large regions of stability of the intersubband superconducting state (ISP) as well as the intrasubband paired phase. In the latter we report almost equal superconducting gaps in the two subbands (equal subband pairing -- ESP), $\Delta_{11}\approx \Delta_{22}$ in spite of the non-zero value of the $\Delta E$ parameter. It is due to the relatively high value of the chemical potential with respect to $\Delta E$ taken in the calculations. At the transition between intra-subband and inter-subband paired phases the superconducting gap is enhanced. It should be noted that for the considered system there is a strong detrimental influence of the intrasubband pairing on the intersubband pair formation. That is why there is no region of coexistence of the two phases in the diagram. This fact can be understood based on Eq.~(\ref{eq:gamma}) whose explicit form is given by
\begin{equation}
\begin{split}
\Gamma_{11(22)}&=V\Delta_{11(22)}+V_{12}\Delta_{22(11)}, \\
\Gamma_{12(21)}&=V_{12}\Delta_{12(21)}+V_{12}\Delta_{21(12)}.
\label{eq:gamma2}
\end{split}
\end{equation}
The terms with $V_{12}\Delta_{12(21)}$ and $V_{12}\Delta_{21(12)}$ correspond to the intersubband pairing, while $V_{12}\Delta_{22(11)}$ refers to the intersubband pair hopping. The latter is operative only when the intrasubband pairs are created ($\Delta_{11}\neq 0$ and $\Delta_{22}\neq 0$). In such a case and when the symmetry of the Cooper pairs tunneling rate between the bands is lifted due to $\Delta E \ne 0$, the $V_{12}\Delta_{22(11)}$ term enhances the disproportion between the electron concentrations in the two subbands. This in turn strongly suppresses the intersubband pairing both in the form with zero total momentum of the Cooper pairs (ISP) and in the FF state [see Fig.~\ref{fig3}(b)]. When the intersubband pair hopping is neglected, such strong competition does not occur and the coexistence region of ESP and ISP appears as shown in Ref.~\onlinecite{Moreo2009}. 

A detrimental impact of the intrasubband pairing on the IFF phase is clearly demonstrated in Fig.~\ref{fig3}(b) which presents the results of calculations with the non-zero total momentum paring included. The range of parameters $(V_{12},V)$ presented in Fig.~\ref{fig3}(b) is marked by the white rectangle in the panel (a). As we can see, in the narrow range of $V$ the IFF stability region gradually decreases with increasing $V$ due to their mutual competition described above. Note that the step-like character of the IFF stability region in Fig.~\ref{fig3}(b) is a result of the relatively small resolution used in our calculations. It is due to the high computional costs required for the calculations with non-zero momentum pairing.

The influence of magnetic field for $V=0.05$ is presented in Fig. \ref{fig4}. It should be noted that in the results presented so far all the paired states correspond to a spin-singlet, subband-triplet pairing with $\Delta_{ST}\neq 0$ and $\Delta_{TS}= 0$, where
\begin{equation}
\begin{split}
 \Delta_{ST} &= \frac{1}{\sqrt{2}}(\langle \hat{c}^{\dagger}_{1\uparrow} \hat{c}^{\dagger}_{2\downarrow}\rangle-\langle \hat{c}^{\dagger}_{2\uparrow} \hat{c}^{\dagger}_{1\downarrow}\rangle),\\
 \Delta_{TS} &= \frac{1}{\sqrt{2}}(\langle \hat{c}^{\dagger}_{1\uparrow} \hat{c}^{\dagger}_{2\downarrow}\rangle+\langle \hat{c}^{\dagger}_{2\uparrow} \hat{c}^{\dagger}_{1\downarrow}\rangle).
 \end{split}
\end{equation}
However, in the presence of external magnetic field, the Zeeman spin-splitting may lead to $|\langle \hat{c}^{\dagger}_{1\uparrow} \hat{c}^{\dagger}_{2\downarrow}\rangle|\neq |\langle \hat{c}^{\dagger}_{2\uparrow} \hat{c}^{\dagger}_{1\downarrow} \rangle|$. This is caused by the fact that the two spin-subbands corresponding to $|2,\uparrow \rangle $ and $|1,\downarrow \rangle$ are closer to each other than the two corresponding to $|1,\uparrow \rangle$ and $|2,\downarrow \rangle$. This mechanism is schematically presented in Fig. \ref{fig4}(e). As a result the spin-singlet, subband-triplet paired state is mixed with spin-triplet, subband-singlet for which $\Delta_{ST}\neq 0$ and $\Delta_{TS}\neq 0$ (ISP$^*$ in the phase diagram, Fig.~\ref{fig4}(a)). The region of stability of such an exotic state appears for relatively large magnetic fields as seen in Fig.~\ref{fig4} (b) and (c), however,  the $\Delta_{TS}$ component of the pairing amplitude is significantly smaller than $\Delta_{TS}$ [Fig.\ref{fig4}(d)]. Note that the ISP$^*$ phase leads to a re-entrance of superconductivity with increasing field $B$ in the range $0.06 \lesssim V_{12}\lesssim 0.1$, where the intersubband paired phase is formed after the intrasubband phase, already suppressed by the Zeeman-splitting. 
\begin{figure}[ht]
\begin{center}
\includegraphics[scale=0.18]{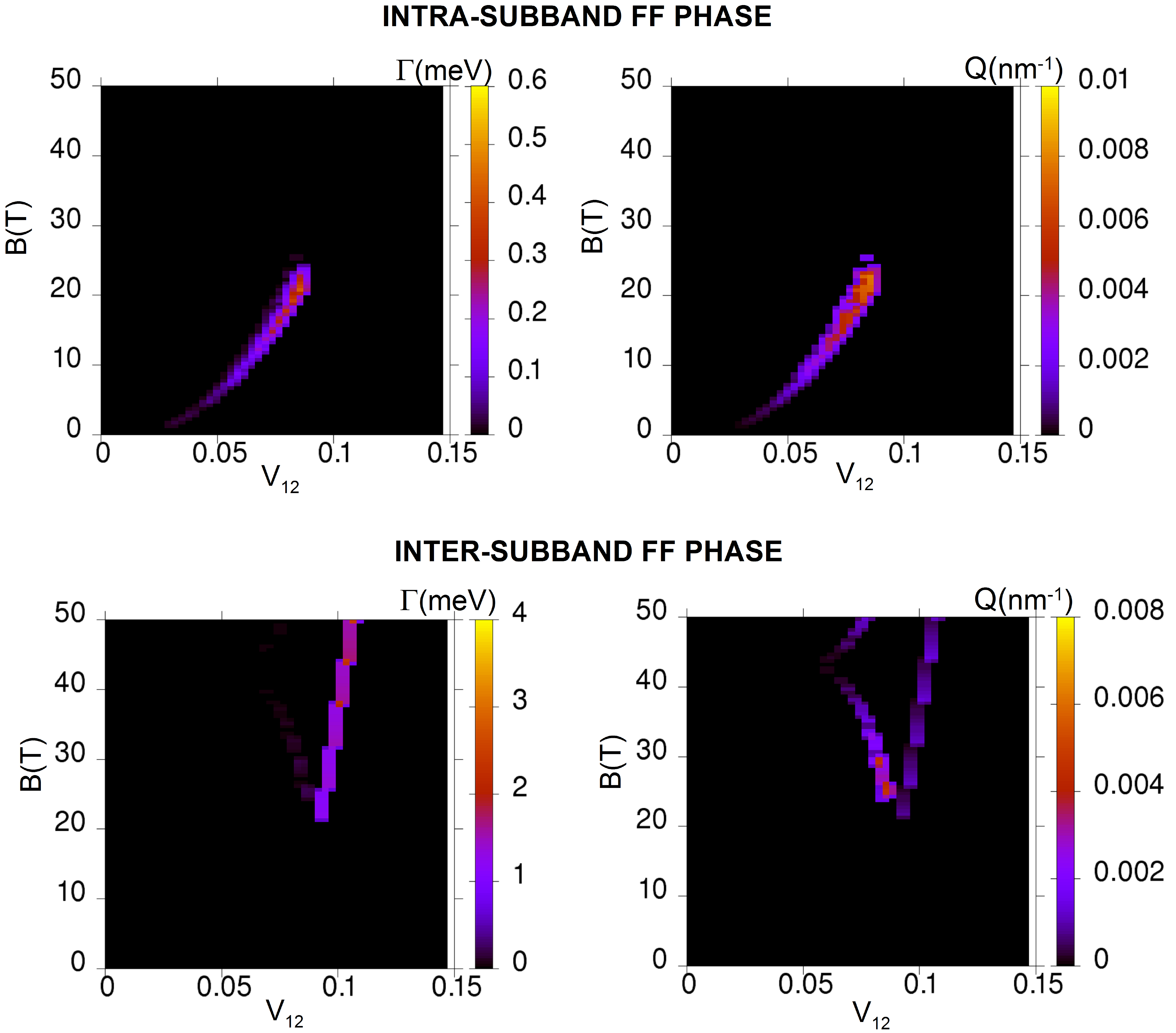}
\caption{The superconducting gaps corresponding to the non-zero momentum pairing within the subbands (a) and between the subbands (b) as functions of $V_{12}$ and $B$. In (c) and (d) the values of the Cooper pair momentum which minimize the system energy are provided.}
\label{fig5}
\end{center}
\end{figure}

Our calculations with $\mathbf{Q}\ne 0$ in the presence of external magnetic field show that the Fulde-Ferrell phase can appear both in the inter- and intra-subband form. The latter is created due to the compensation of the Fermi wave vector mismatch caused by the Zeeman splitting within the two bands (FF phase in Fig.~\ref{fig4}(a)). As presented in Fig.~\ref{fig1}, in general the $\mathbf{Q}$ vector for each of the subbands and between them could be different when $H\ne 0$. However, due to the fact that $\Delta E \ll \mu$ the wave vector mismatch induced by the magnetic field at the Fermi level is nearly the same for both the subbands, and therefore we assume $\mathbf{Q_1}=\mathbf{Q_2}$. The corresponding regions of stability of both phases are shown in Fig. \ref{fig5} where the amplitude for the non-zero pairing between the subbands and within the subbands is plotted as a function of both $V_{12}$ and $B$. Additionally, in (c) and (d) the values of the Cooper pair momentum which minimize the system energy are provided. Note that the ordinary FF phase induced by the Zeeman effect appears only at the border between the ESP phase and NS state. In contrast to that, the IFF phase emerges at both the borders NS/ISP$^*$ and ISP$^*$/ISP.

\section{Summary and outlook}
\label{sec:4}
In the present paper, we have analyzed the intersubband pairing in free-standing Pb nanofilms where the subbands are created due to the quantization effect induced by the confinement in the growth direction. In order to determine the principal features of the paired state we have used a model consisting of two parabolic subbands separated by the energy $\Delta E$. Non-zero values of the latter is detrimental for the intersubband pairing as it generates a Fermi wave vector mismatch between the particles with opposite spins and momenta. However, as we show here such a mismatch between the bands can be compensated by the non-zero total momentum of the Cooper pairs which leads to a appearance of the spontaneous Fulde-Ferrell intersubband state. The interesting feature of such state is that it can be formed without the necessity of applying any external magnetic or electric field. Since the inter-subband pair hopping processes together with the intra-subband pairing enhance the disproportion between the electron concentrations in the two bands, the coexistence of the inter- and intra-subband paired states does not appear. The same mechanism leads to suppression of the FF stability region with increasing intrasubband paring. 

Our calculations have shown that in relatively high external magnetic field, due to the Zeeman splitting of the bands, an exotic inter-band paired state appears, for which the spin-singlet, subband-triplet as well as spin-triplet, subband-triplet Cooper pairs are created. The appearance of such a phase has a re-entrant character with increasing magnetic field after the intra-subband paired state is destroyed. The spontaneous inter-subband Fulde-Ferrell phase appears in all of the analyzed situations close to the transition between the ISP phase and other states appearing in the phase diagram such as the normal state or the intra-subband paired state.

Although in our study we consider a particular case of Pb metallic nanofilm, the appearance of a spontaneous Fulde-Ferrel state could be energetically favorable for any multiband superconductor such as MgB$_2$\cite{Souma2003,Giubileo2001,Chen2001,Szabo2001} or iron-based superconductors\cite{Kuroki2008, Dai2008, Jeglic2010, Singh2008,Dai2008,Zegrodnik2014}. In the latter, the bottoms of the bands between, which the pairing may appear coincide, however, the Fermi wave vector mismatch appears due to a specific electronic structure. In general, the Fermi wave vector mismatch has to be relatively small for the IFF phase to appear, in systems with weak intersubband coupling constant. This could be seen in our calculations where the non-zero-momentum pairing emerged even for low $V_{12}$ if $\Delta E$ was small. We propose that those conditions could be also satisfied in the two-dimensional superconducting electron gas created at LaAlO$_3$/SrTiO$_3$ interfaces\cite{Joshua2012, Reyren2007, Scheurer2015, Trevisan2018} where the appearance of the intersubband pairing is confirmed experimentally\cite{Singh2018} and can be controlled by gating (doping)\cite{Trevisan2018}. Also the proposed state could appear for a system consisting of two types of particles with two different masses between which a pairing mechanism is realized. Such a scenario can be realized in a ultracold Fermi gas trapped in an optical lattice Note that, since the existence of the IFF phase does not require any magnetic field, its appearance is independent of the Maki criterion for the Cooper pair decomposition, which plays an important role for the case of the conventional FF state. This significantly extends the range o materials in which this phase can possibly be observed.

\section{Acknowledgement}
This work was supported by National Science Centre, Poland (NCN) according to decision 2017/26/D/ST3/00109 and in part by PL-Grid Infrastructure.                                                       

%

\end{document}